\documentclass[aps,prb,twocolumn,showpacs,showkeys,preprintnumbers,groupedaddress,amsmath,amssymb,]{revtex4-1}

\usepackage{graphicx}
\usepackage{dcolumn}
\usepackage{bm}
\usepackage{hyperref}
\usepackage[dvips]{color}
\usepackage{booktabs}
\usepackage{longtable}
\usepackage{array}
\usepackage{dcolumn,booktabs}

\begin{document}

\title{Temporal universal conductance fluctuations in RuO$_2$ nanowires due to mobile defects}

\author{An-Shao Lien$^1$}
\author{L. Y. Wang$^2$}
\email{luyao.ep88g@nctu.edu.tw}
\author{C. S. Chu$^2$}
\author{Juhn-Jong Lin$^{1,2,}$}
\email{jjlin@mail.nctu.edu.tw}

\affiliation{$^1$Institute of Physics, National Chiao Tung University, Hsinchu 30010, Taiwan\\
$^2$Department of Electrophysics, National Chiao Tung University, Hsinchu 30010, Taiwan}

\begin{abstract}

Temporal universal conductance fluctuations (TUCF) are observed in RuO$_2$ nanowires at cryogenic temperatures. The fluctuations persist up to very high $T \sim 10$ K. Their root-mean-square magnitudes increase with decreasing $T$, reaching $\sim 0.2 e^2/h$ at $T \lesssim 2$ K. These fluctuations are shown to originate from scattering of conduction electrons with rich amounts of mobile defects in artificially synthesized metal oxide nanowires. TUCF characteristics in both one-dimensional saturated and unsaturated regimes are identified and explained in terms of current theories. Furthermore, the TUCF as a probe for the characteristic time scales of the mobile defects (two-level systems) are discussed.

\end{abstract}


\maketitle

\section{Introduction}

The study of quantum-interference effects (QIE) is a central theme in the electron transport properties of mesoscopic and nanoscale conductors.
\cite{mesoscopic} When the dimensions of a miniature conductor become comparable to the electron dephasing length $L_\varphi$, the QIE cause notable
corrections to the classical (magneto)conductivity, and the Ohm's law may no longer be valid. $L_\varphi$ is the characteristic length scale over which the
electron wavefunction maintains its deterministic phase memory. \cite{Lin-jpcm02}

One of the most extensively studied phase-coherence phenomena is the universal conductance fluctuations (UCF). \cite{Washburn92} UCF has the unique feature
that their fluctuation magnitudes increase with the lowering of $T$ (usually, at temperatures $< 1$ K). \cite{LeePA85,Altshuler85,Altshuler85b,Lee-prb87} In most cases, the properties of UCF are studied by measuring the conductance $G$ as a function of magnetic field $B$ or Fermi energy, where the UCF exhibit reproducible but aperiodic ``fingerprints," provided that the device is constantly maintained at low $T$. On the other hand, the temporal UCF (TUCF) have not been widely seen in experiments, where $G$ fluctuates with time. \cite{Feng,Feng2,ng91} Previously, ``direct" observations of TUCF in nonmagnetic metals had only been made in thin Bi wires and films \cite{Beutler} and Ag films \cite{Meisenheimer} at very low $T \lesssim 0.5$ K. \cite{spin-glass} Besides, the power spectra as a function of $T$ or $B$  had been studied, \cite{Birge89,Birge90,Trionfi07} which effectively had integrated over the TUCF events over time. These phenomena were ascribed to perpetual fluctuations of single scattering centers between their bistable positions with time scales $\tau_d$, which values have a very broad distribution. \cite{Birge90} The current theory on TUCF has made no connection with $\tau_d$, \cite{LeePA85,Altshuler85,Altshuler85b,Lee-prb87,Feng,Feng2} though the switching rates of particular defects have been measured in several cases. \cite{Birge03,Chun96}

Ruthenium dioxide (RuO$_2$) crystallizes in the rutile structure and exhibits metallic conductivities, which can be described by the standard Boltzmann equation. \cite{Glassford94} Quasi-one-dimensional (1D) RuO$_2$ nanowires (NWs) are stable in the ambient environment and could be applied as interconnects in nanoelectronic devices. \cite{LinYH-nano08} Previously, the magnetoresistance (MR) in the weak-localization (WL) effect had been studied in three-dimensional (3D) RuO$_2$ thick films. \cite{Lin-prb99} In that case, the UCF were not observed due to the macroscopic sample dimensions. In sharp contrast, we find in this work that the $G$ of individual RuO$_2$ NW fluctuates markedly with time. Thus, the MR is largely smeared out and difficult to trace. The TUCF persist up to a very high $T \sim 10$ K. Furthermore, our measured TUCF can be quantitatively described by the saturated and unsaturated theories of Feng, \cite{Feng2} which have been predicted for two decades but not yet been experimentally tested for 1D. We thus identify the microscopic origin for our TUCF to be the existence of rich amounts of mobile defects in metal oxide NWs. The dynamic defects could arise from oxygen nonstoichiometries. \cite{Chen04}

This paper is organized as follows. Section II contains our experimental method. Section III contains our experimental results and theoretical analysis. A proposal for using the TUCF as a sensitive probe for the characteristic time scales of the mobile defects is also discussed. Our conclusion is given in Sec. IV.

\section{Experimental Method}

Single-crystalline RuO$_2$ NWs were grown by the thermal evaporation method. The morphology and atomic structure of the NWs were studied by the scanning
electron microscopy (SEM) and the transmission electron microscopy. \cite{LiuYL-apl07} Four-probe individual NW devices were fabricated by the electron-beam lithography. \cite{LinYH-nano08} [Figure \ref{fig1}(c) shows an SEM image of the NW67 nanowire device.] The resistance measurements were performed on standard $^4$He and $^3$He cryostats. \cite{Hsu10} A Linear Research LR-700 ac resistance bridge operating at a frequency of 16 Hz was employed for resistance measurements. An excitation current of $\lesssim 100$ nA (so that the voltage drop $\lesssim k_BT/e$, where $k_B$ is the Boltzmann constant) was applied to avoid electron heating. Table 1 lists the parameters of the four NWs studied in this work. The samples are named according to their diameter.

\begin{table}
\caption{Parameters for RuO$_2$ NWs. $d$ is the diameter, $L$ is the voltage probe distance in a four-probe geometry, $D$ is the diffusion constant, and
$\ell$ is the electron mean free path.}

\begin{ruledtabular}
\begin{tabular}[t]{lcccccc}

Nanowire & $d$ & $L$ & $\rho$(300 K) & $\rho$(10 K) & $D$ & $\ell$ \\

 & (nm) & ($\mu$m) & ($\mu\Omega$ cm) & ($\mu\Omega$ cm) & (cm$^2$/s)
 & (nm) \\ \hline

NW67  & 67  & 1.5  & 180 & 125 & 3.5  & 1.6\\
NW120 & 120 & 0.73 & 200 & 165 & 2.6 & 1.2\\
NW54  & 54  & 0.69 & 580 & 450 & 0.95 & 0.44\\
NW47  & 47  & 1.0  & 780 & 616 & 0.71 & 0.33\\

\end{tabular}
\end{ruledtabular}
\end{table}

\section{Results and Discussion}
\subsection{Experimental results and comparison with theory}

\begin{figure}
\includegraphics[width=8.0cm, height=5.0cm]{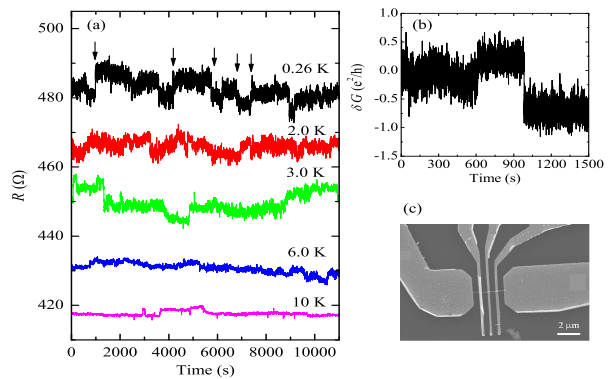}
\caption{(color online). Resistance/conductance fluctuations in NW67 nanowire. (a) Resistance versus time at five temperatures, as indicated. The arrows indicate a few slow jumps. Data for 6.0, 3.0, 2.0, and 0.26 K are offset for 10, 20, 30, and 40 $\Omega$, respectively, for clarity. (b) Conductance variation $\delta G = G - \langle G \rangle$ versus time at 0.26 K. (c) An SEM image of this NW device. \label{fig1}}
\end{figure}

Figure \ref{fig1}(a) shows the resistance $R$ as a function of time for the NW67 nanowire at five $T$ values. The fluctuations in $R$ with time is evident and can be categorized into two types: the first type characterizes fast fluctuations, and the second type characterizes a few discrete slow fluctuations. \cite{drift} In the case of slow fluctuations (jumps), indicated by arrows for some of them, $R$ changes abruptly from one average value to another. The time interval between such consecutive jumps is irregular and relatively long. Fast fluctuations, on the other hand, involve many nearly simultaneous jumps with overlapping in their time intervals. These fluctuations are characterized by a broad distribution of time scales and a range of fluctuation magnitudes. \cite{ng91} These jumps can not be individually resolved by the usual electrical-transport measurement technique. (In this experiment, 5 data points were acquired per second.) Most notably, the magnitudes of the fast fluctuations are greater at lower $T$. This is a pivotal signature of the UCF mechanism. As $T$ increases, the fast fluctuations decrease monotonically and diminish around $\sim 10$ K, where the $R$ fluctuation magnitudes level off to the instrumental noise level ($\lesssim 1$ $\Omega$).

We plot in Fig.~\ref{fig1}(b) the temporal variation of the conductance fluctuation $\delta G = G - \langle G \rangle$ of the NW67 nanowire at 0.26 K. In units of the quantum conductance $e^2/h$, $\langle G \rangle$ is the measured conductance averaged over time in Fig.~\ref{fig1}(a). The fast fluctuations have a ``peak-to-peak" magnitude of $\delta G \approx 0.5 e^2/h$. Also presented are two slow jumps occurring at $\approx$ 610 s, 980 s on the time scale, and with abrupt $G$ changes of $\approx 0.2 e^2/h$, $-0.7 e^2/h$, respectively. Therefore, the typical magnitudes of slow jumps at low $T$ are also of a fraction of $e^2/h$. Because the fast fluctuations provide immense events for a reliable statistical analysis, a critical check on the theory of mobile-defect-induced UCF predicted by Feng \cite{Feng2} becomes possible. We shall focus on the fast fluctuations in the rest of this paper. As for the slow fluctuations, they had previously been ascribed to the UCF mechanism, \cite{Beutler,Meisenheimer} but it is much less eventful here as to warrant a quantitative analysis.

\begin{figure}
\includegraphics[width=8.0cm, height=4.5cm]{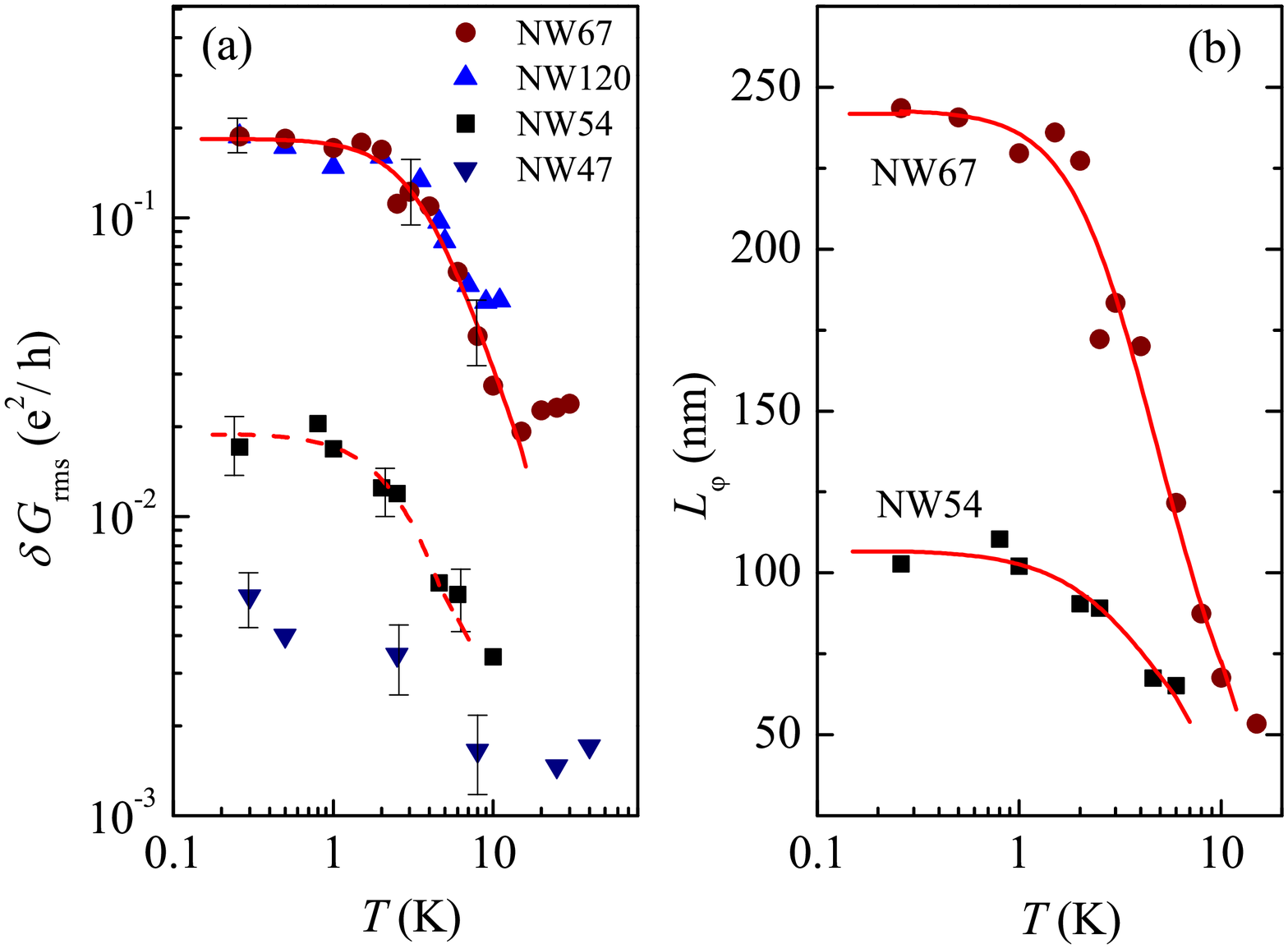}
\caption{(color online). (a) log$_{10}$($\delta G_{\rm rms}$) as a function of log$_{10}$($T$) for four RuO$_2$ NWs. The solid (dashed) curve drawn through NW67 (NW54) nanowire is a least-squares fit to Eq. (\ref{Q1DG}) [(Eq. \ref{Q1DG-un})]. Note that the system noise has not been subtracted from these results. (b) $L_\varphi$ as a function of log$_{10}$($T$) for NW67 and NW54 nanowires. The solid curves are least-squares fits to Eq. (\ref{eq3}). \label{fig2}}
\end{figure}

To quantify the fast UCF, we start with the root-mean-square magnitude of the conductance fluctuation $\delta G_{\rm rms} = \sqrt{ \langle ( G - \langle G
\rangle )^2 \rangle }$. Here $\langle ... \rangle$ denotes averaging over a proper time interval while excluding the slow fluctuations. Figure \ref{fig2}(a) shows $\delta G_{\rm rms}$ as a function of $T$ in double-logarithmic scales for the four NWs listed in Table 1. This figure reveals that $\delta G_{\rm rms}$ increases with decreasing $T$ in every NW until below about 1--2 K (depending on sample), where $\delta G_{\rm rms}$ tends to saturate. (The level off at $T > 10$ K is due to the background noise.) While such $T$ behavior of $\delta G_{\rm rms}$ is similar for all samples, Fig.~\ref{fig2}(a) clearly indicates that the $\delta G_{\rm rms}$ magnitude varies greatly from one NW to another. The magnitudes of $\delta G_{\rm rms}$ in the NW67 and NW120 nanowires are more than one order of magnitude higher than that in the NW54 nanowire which, in turn, is a factor of $\sim 5$ larger than that in the NW47 nanowire. Such sensitive sample dependent $\delta G_{\rm rms}$ magnitude provides a strong evidence that our measured $\delta G_{\rm rms}$ must originate from specific NW properties rather than from instrumental electronics. Note that as $T$ decreases from 10 to 0.26 K, the magnitude of $\delta G_{\rm rms}$ in the NW67 nanowire increases by about one order of magnitude, from $\approx 0.02 e^2/h$ to $\approx 0.2 e^2/h$. In contrast, the TUCF in submicron Ag films studied by Giordano and coworkers could only be seen at much lower $T \lesssim 0.5$ K.  \cite{Meisenheimer} This strongly suggests that dynamic defects in RuO$_2$ NWs are markedly more numerous and/or vigorous than those in conventional lithographic metal mesoscopic structures.

Theoretically, the UCF in different dimensionalities and under different conditions have been studied by Altshuler, \cite{Altshuler85} and Lee and coworkers. \cite{Lee-prb87} Altshuler and Spivak, \cite{Altshuler85b} and Feng, Lee and Stone \cite{Feng} proposed that the UCF are very sensitive even to the motion of single or a few scattering centers. Explicitly, Feng \cite{Feng2} predicted that, in a 1D rectangular wire with transverse dimensions $L_x$ and $L_y$ and longitudinal dimension $L_z$ ($> L_\varphi$), the fluctuation magnitude in the ``saturated" regime is given by
\begin{equation}
{(\delta G_{\rm rms}) ^2 } = 177.7 \langle G^2 \rangle \frac{1}{{k_F ^4 \ell^2
}}\frac{{L_\varphi ^3}}{{L_x ^2 L_y ^2 L_z}} \,, \label{Q1DG}
\end{equation}
where $k_F$ is the Fermi wavenumber, and $\ell$ is the electron mean free path. A sample falls in the saturated regime if $\beta$, the ratio of the number of mobile defects to the number of total (static and mobile) defects, satisfies the condition $\beta \gg (\ell/L_\varphi)^2$. Conversely, in the ``unsaturated" regime $\beta \ll (\ell/L_\varphi)^2$, Feng predicted
\cite{Feng2}
\begin{equation}
{(\delta G_{\rm rms}) ^2 } = 27 \pi^2 \beta \bar{C} \langle G^2 \rangle
\frac{1}{{k_F ^4 \ell^4 }}\frac{{L_\varphi ^5}}{{L_x ^2 L_y ^2 L_z}} \,,
\label{Q1DG-un}
\end{equation}
where $\bar{C}$ is a constant (typically, $\sim 0.1$). The predictions of Eqs. (\ref{Q1DG}) and (\ref{Q1DG-un}) have thus far not been experimentally tested.
(The corresponding 2D regime have previously been studied by Birge and coworkers. \cite{Birge90})

For the NW67 nanowire, $k_F \ell \approx 13$. Thus, the NW is weakly disordered and satisfies the prerequisite condition of the UCF mechanism, namely,
$G > e^2/h$. The solid curve through the NW67 nanowire in Fig.~\ref{fig2}(a) was obtained by least-squares fitting the data to Eq. (\ref{Q1DG}), with $L_\varphi$
being the sole adjustable parameter. We took $L_x = L_y = d$ and $L_z = L$, the NW segment between the two voltage probes. The extracted $L_\varphi$ as a
function of log$_{10}$($T$), plotted in Fig.~\ref{fig2}(b), is subject to further scrutiny below. Together we see that Eq. (\ref{Q1DG}) well describes the data. Furthermore, $L_\varphi$ decreases monotonically from 245 nm at 0.26 K to 73 nm at 10 K. This provides an assuring check that this NW is 1D ($L_\varphi > d$) for the UCF effect. Since $L_\varphi \gg \ell$ in our NWs, the electrons undergo diffusive motion such that the UCF physics rather than the ``local interference" mechanism ($\ell \gg L_\varphi$) \cite{Feng2,ng91} is responsible for the $G$ fluctuations we observed in this work.

The physical meaning of $L_\varphi (T)$ is further examined below. In general, $L_\varphi$ can be written as \cite{Lin-jpcm02}
\begin{equation}
\frac{1}{L_\varphi^2 (T)} = \frac{1}{L_0^2} + \frac{1}{L_{ee}^2 (T)} + \frac{1}{L_{ep}^2 (T)}  \,, \label{eq3}
\end{equation}
where $L_0$ is a constant, whose origins are a subject of elaborate investigations. \cite{Lin-jpcm02,Lin-prb87b,Mohanty97,Pierre03,Huang-prl07} The 1D electron-electron ($e$-$e$) dephasing length is $L_{ee} = \sqrt{D \tau_{ee}}$, where $D$ is the diffusion constant, and $1/\tau_{ee} = A_{ee} T^{2/3}$. \cite{Lin-jpcm02,Pierre03,Altshuler82} The electron-phonon dephasing length is $L_{ep} = \sqrt{D \tau_{ep}}$, where $1/\tau_{ep} = A_{ep} T^p$. The value of the exponent $p$ depends on the $T$ interval and the degree of disorder in the sample (typically, $2 \leq p \leq 4$). \cite{Lin-jpcm02,Sergeev-prb00} Figure \ref{fig2}(b) shows that our extracted $L_\varphi$ in the NW67 nanowire can be well described by Eq. (\ref{eq3}). That the fitting parameters, listed in Table II, have the appropriate orders of magnitude lends strong support to this finding. \cite{e-ph} For the parameter $A_{ee}$, the theoretical prediction \cite{Lin-jpcm02,Pierre03,Altshuler82} $(A_{ee})^{\rm th} = [(e^2\sqrt{D}Rk_B) / (2\sqrt{2} \hbar^2 L)]^{2/3}$ = $1.6 \times 10^9$ K$^{-2/3}$ s$^{-1}$ is in good agreement with the experimental value. Furthermore, our measured UCF have a saturated value $\delta G_{\rm rms}(0.26 \, {\rm K}) \approx 0.2 e^2/h$, which is smaller than that of a  mesoscopic 1D sample, $0.73 e^2/h$. \cite{LeePA85,Lee-prb87} This can be understood in light of the fact that our NW length $L \approx 6 L_\varphi (0.26\, {\rm K}) \approx 6 L_0$. The small ensemble of subsystems leads to a reduction factor $\sim 1/\sqrt{6} \approx 0.4$, which is in very good consistency with the measured value.

For the NW54 nanowire, the $\delta G_{\rm rms}$ magnitude is one order of magnitude smaller than that in the NW67 nanowire. We find out that the measured $\delta G_{\rm rms}$ can be least-squares-fitted with Eq. (\ref{Q1DG-un}) [the dashed curve in Fig.~\ref{fig2}(a)]. This result implies that the NW is in the unsaturated regime, with the fraction of mobile defects $\beta$ much smaller than that in the NW67 nanowire. Numerically, we obtained $\beta \approx 1 \times 10^{-6}$, which corresponds to a dynamic impurity concentration $\beta / \ell^3 \sim 1 \times 10^{16}$ cm$^{-3}$ or $\sim 3$ mobile defects in a phase-coherence segment. This result illustrates the extreme sensitivity of UCF to a few mobile defects. In contrast, $\beta$ for the NW67 nanowire is estimated, with $\beta > (\ell/L_\varphi)^2 \sim 10^{-4}$, to be more than one order of magnitude higher. \cite{TLS} A confirmation of the unsaturated regime for the NW54 nanowire is obtained from Fig.~\ref{fig2}(b), when $L_\varphi$ varies from 105 nm (0.26 K) to 65 nm (6 K). With $\ell \approx 0.44$ nm (Table I), the criterion $\beta \ll (\ell/L_\varphi)^2$ is satisfied. Furthermore, our extracted value of $A_{ee}$ (Table II) is in good agreement with the theoretical value $(A_{ee})^{\rm th} = 3.1 \times 10^9$ K$^{-2/3}$ s$^{-1}$. (No fitting is done for the NW47 nanowire due to the small $\delta G_{\rm rms}$.)

\begin{table}
\caption{Fitting parameters for $L_\varphi (T)$ defined in Eq. (\ref{eq3}). $A_{ee}$ is in K$^{-2/3}$ s$^{-1}$, $A_{ep}$ in K$^{-p}$ s$^{-1}$, and $L_0$ in nm.}

\begin{ruledtabular}
\begin{tabular}[t]{lcccc}

Nanowire & $A_{ee}$ & $A_{ep}$ & $p$ & $L_0$ \\ \hline
NW67 & $3.0 \times 10^9$ & $3.7 \times 10^8$ & 2.1$\pm$0.2 & 240 \\
NW54 & $6.0 \times 10^9$ & $1.3 \times 10^9$ & 1.9$\pm$0.2 & 105 \\

\end{tabular}
\end{ruledtabular}
\end{table}

\begin{figure}
\includegraphics[width=8.0cm, height=5.0cm]{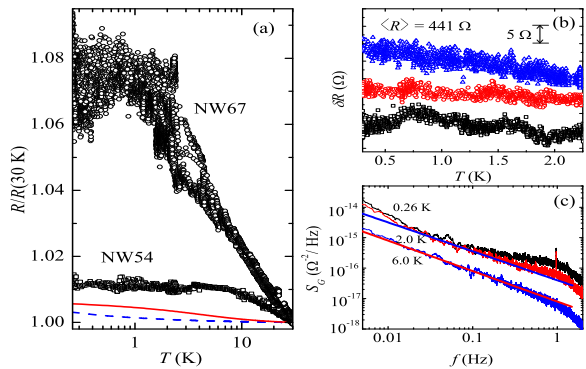}
\caption{(color online). (a) Variation of $R/R$(30\,K) with log$_{10}$($T$) of NW67 and NW54 nanowires. The solid (dashed) curve indicates 1D WL (EEI) contribution with the sample parameters of NW67 nanowire [that for NW54 (not shown) are even smaller]. (b) $\delta R$ versus $T$ of NW67 nanowire from three cooldowns. Data are offset for clarity. (c) Noise power spectrum log$_{10}$($S_G$) as a function of log$_{10}$($f$) of NW67 nanowire. The solid line through the 2.0 (6.0) K data is a least-square fit to $f^{-0.95}$ ($f^{-1.0}$) frequency dependence. \label{fig3}}
\end{figure}

Our assertion that the observed TUCF are due to mobile defects is further supported by examining the $T$ dependence of $R$. Figure \ref{fig3}(a) shows the normalized resistance, $R/R$(30\,K), as a function of $T$ for the NW67 and NW54 nanowires from several cooldowns. In the NW67 nanowire, $R$ increases by a large amount $\approx (7 \pm 2)$\% as $T$ is reduced to 0.26 K. The $R$ increases due to 1D WL and $e$-$e$ interaction (EEI) effects predict an increase $< 1$\% in this sample [the solid and dashed curves in Fig. \ref{fig3}(a), respectively]. The discrepancy demands an extra, governing contribution to the observed ``large" low-$T$ resistance rise. (Similar conclusion applies to the NW54 nanowire.) In contrast, such a discrepancy was not found in conventionally deposited Bi wires. \cite{Beutler} This extra contribution originates most likely from the scattering of conduction electrons with mobile defects (two-level systems, TLS). \cite{Anderson72} Indeed, a logarithmic $T$ dependence followed by a ``saturation" behavior is consistent with the TLS-induced $R$ rise. \cite{Huang-prl07,Zawa98,Cichorek03} Moreover, Fig. \ref{fig3}(a) indicates that, as $T$ reduces, the measured $R$ distributes over a wider range of $R$ values at a given $T$. This is a direct manifestation of the TUCF. Note that both the increments in $R$ and in the $R$ distribution are markedly larger for the NW67 than for the NW54 nanowire, suggesting again a much higher level of mobile defects in the former NW. Effect of thermal cycle is shown in Fig.~ \ref{fig3}(b) for the NW67 nanowire from three cooldowns. The detailed variations in $R$ differ in the three runs, obviously due to the rearrangement of mobile defects (while the peak-to-peak fluctuation magnitudes remain similar, $\sim 5$ $\Omega$ or $\sim 0.5 e^2/h$). In addition, our power spectral analysis of the fast fluctuation data shown in Fig.~\ref{fig1}(a) reveals a $1/f^\alpha$ frequency dependence with $\alpha \approx 1.0 \pm 0.1$ in the interval $\sim$ 0.005--1 Hz [Fig.~\ref{fig3}(c)], as predicted by the UCF mechanism. \cite{Feng,Feng2}

\subsection{Time scales of mobile defects}

We would like to point out that TUCF could provide more microscopic information about the mobile defects than one previously thought. Specifically, we are referring to the time scale $\tau_d$ in which the mobile defects perpetually fluctuate between their bistable positions. Though the current TUCF theory \cite{LeePA85,Altshuler85,Altshuler85b,Lee-prb87,Feng,Feng2} has left this $\tau_d$ aspect unaddressed, it does lead one to a vital piece of information, namely, the number of mobile defects $N$ in a phase-coherence segment. In this work, we have obtained $N \gtrsim 30$ ($\sim 3$) for the NW67 (NW54) nanowire. These mobile defects can be configured in $N_{\rm c} \simeq 2^N$ possible ways where their respective $G$ values vary with $\delta G_{\rm rms} \sim e^2/h$, according to UCF. However, the TUCF would not have been observed in an experiment with a measuring time $T_{\rm m}$ for each taking of the $G$ data, if we were in the regime $T_{\rm c} \ll T_{\rm m}$, where $T_{\rm c}$ is the time for the mobile defects to evolve through the entire $N_c$ configurations. Thus the condition $T_{\rm c} \gtrsim T_{\rm m}$ must hold in this work.

Assuming that $T_{\rm c} \sim N_{\rm c} \tau_d$, the condition that the dynamics of these $N$ mobile defects can be observed in our TUCF experiment provides us a lower bound to the $\tau_d$ of these defects, namely, $\tau_d \gtrsim \tau_{d,{\rm min}} \equiv T_{\rm m}/N_{\rm c}$. Specifically, taking $T_{\rm m} \sim 0.1$ s, we have $\tau_{d,{\rm min}} \sim 0.1$ ns for $N \sim 30$ (NW67 nanowire), and $\tau_{d,{\rm min}} \sim 0.01$ s for $N \sim 3$ (NW54 nanowire). This large variation in $\tau_{d,{\rm min}}$ or, for that matter, $\tau_d$, values of the mobile defects is consistent with current understanding. The value of $\tau_d \sim 0.1$ ns is in quantitative accord with the properties of TLS in disordered metals. \cite{Birge90,Golding78} $T_{\rm m}$ is the upper bound of $\tau_d$, since the $\tau_d \gg T_{\rm m}$ cases are the slow fluctuation events, which has been excluded in our $\delta G_{\rm rms}$ analysis.  We also note that the above discussion does not rule out the possible existence of very fast TLS with time scales shorter than $\tau_{d,{\rm min}}$ in our nanowires. Our results here thus imply a possible use of $T_{\rm m}$ for a more microscopic probe of the mobile  defects.

\section{Conclusions}

We have observed temporal universal conductance fluctuations in RuO$_2$ nanowires up to very high temperatures of $\sim 10$ K. The TUCF originate from rich and vigorous mobile defects (e.g., oxygen nonstoichiometries) in as-grown nanowires. The measured TUCF magnitudes are well described by the 1D theory and the numbers of mobile defects have been evaluated. Furthermore, we discuss that the microscopic information on the characteristic time scales of the mobile defects may be learned from TUCF
measurements.

\begin{acknowledgments}

The authors are grateful to Y. H. Lin and S. P. Chiu for helpful experimental assistance, and B. L. Altshuler, N. Giordano and N. O. Birge for valuable discussion. This work was supported by Taiwan National Science Council through Grant Nos. NSC 99-2120-M-009-001 and NSC 100-2120-M-009-008, and by the MOE ATU Program.

\end{acknowledgments}


\begin{thebibliography}{99}
\bibitem{drift} A small drift is seen, e.g., in the curve at 6.0 K, which is likely due to instrumental electronics and is irrelevant to the TUCF.
\bibitem{e-ph} The two fitted $A_{ep}$ values scale closely with $1/\ell$ (Ref. \onlinecite{Sergeev-prb00}), further supporting our data analysis in this work.
\bibitem{TLS} This estimate is in good accord with the typical TLS concentrations $\sim$ $10^{16}$--$10^{18}$ cm$^{-3}$ found in polycrystalline metals and metallic glasses at 1 K, see Refs. \onlinecite{Birge90} and \onlinecite{Chun96}.
\end{thebibliography}
\end{document}